\documentclass[a4paper]{svproc}
\usepackage{amsmath}
\usepackage{xcolor}
\usepackage{amsfonts}
\usepackage{amsmath}
\usepackage{graphicx}
\usepackage{hyperref}
\usepackage{booktabs}
\usepackage{makecell}
\usepackage{adjustbox}

\usepackage{url}

\begin{document}
\mainmatter              
\title{Performance Benchmarking and Optimisation of Clustering Algorithms for Local and Non-Local Similarity Measure in Medical Image Analysis}
\titlerunning{Optimisation of Clustering Algorithms}  
%
\author{Sisipho Hamlomo\inst{1,2}\and
Marcellin Atemkeng\inst{1,3}}
\authorrunning{Sisipho Hamlomo and Marcellin Atemkeng} 
%
\tocauthor{Sisipho Hamlomo and Marcellin Atemkeng}
\institute{%
Department of Mathematics, Rhodes University, PO Box 94, Makhanda, 6140, South Africa\and
Department of Statistics, Rhodes University, PO Box 94, Makhanda, 6140, South Africa\and
National Institute for Theoretical and Computational Sciences (NITheCS), Stellenbosch 7600, South Africa\\
\email{\{s.hamlomo, m.atemkeng\}@ru.ac.za}}

\maketitle              

\begin{abstract}
Medical imaging generates high-resolution images posing significant storage, transmission, and computational challenges. While low-rank matrix approximation (LoRMA) techniques offer efficient compression by exploiting structural redundancy, global approaches often fail to preserve local details critical for diagnosis. This paper focuses on clustering techniques that exploit non-local self-similarity to identify structurally similar regions in medical images. These clusters can be used for post-processing tasks such as adaptive image compression. We evaluate five clustering techniques: k-means, mini-batch k-means, agglomerative hierarchical clustering, balanced iterative reducing and clustering using hierarchies (BIRCH), and bisecting k-means across MRI, ultrasound, and chest X-ray modalities. All clustering techniques were optimised using random search, and cluster quality was assessed using the Silhouette score, the Davies-Bouldin (DB) index, and the Calinski-Harabasz (CH) index. Results demonstrate that standard k-means and bisecting k-means generally achieve strong cluster cohesion and separation across modalities. However, they tend to form a small number of clusters with high intra-cluster variability, limiting their effectiveness for post-processing tasks such as adaptive compression. Agglomerative clustering outperformed other techniques for MRI and ultrasound in terms of intra-cluster homogeneity, making it more suitable for preserving fine diagnostic details. For chest X-rays, mini-batch k-means achieved the best balance between clustering quality and intra-cluster compactness. BIRCH consistently underperformed across all modalities.
\keywords{Medical image compression, clustering, low-rank matrix approximation, hyperparameter tuning}
\end{abstract}
\section{Introduction}
Medical imaging plays an important role in modern healthcare, enabling accurate diagnosis, treatment planning, and disease monitoring. Advances in imaging modalities such as magnetic resonance imaging (MRI), computed tomography (CT), and ultrasound have significantly enhanced image resolution and diagnostic capability \cite{tempany2001advances,muller2002computed,hussain2022modern}. However, these high-resolution images produce vast amounts of high-dimensional data, posing data storage, transmission, and computational processing challenges during image analysis \cite{lee2017medical,li2018large,panayides2020ai}. In addition, medical images often suffer from noise, artefacts, and variations due to different acquisition conditions, further complicating efficient image analysis \cite{goyal2018noise}. Therefore, efficient image compression becomes essential to reduce the cost burden on healthcare data management systems. To address these challenges, various image compression techniques have been explored, broadly categorised into lossless and lossy methods. While lossless techniques preserve all original data, they often achieve low compression ratios, which limits their practicality in bandwidth-constrained and storage-limited environments \cite{kowalik2015quest,oh2010jpeg2000}. Consequently, lossy compression methods, which aim to reduce data redundancy while preserving and extracting important features, have received increased attention. Standard image compression techniques such as discrete cosine transform (DCT)-based JPEG \cite{watson1994image,more2013jpeg,raid2014jpeg} and wavelet-based JPEG2000 \cite{walker2001wavelet,usevitch2002tutorial,talukder2010haar} have been widely applied. These methods have the advantage of being computationally efficient; however, they often struggle to maintain the fine structural details and diagnostic features inherent in medical images, leading to the potential loss of clinically significant information \cite{oh2002medical}. To address these shortcomings, researchers have explored techniques such as sparsity-based and dictionary learning methods \cite{horev2012adaptive,watkins2018image,anandan2016medical}, which introduce additional complexity and computational demands while improving the fidelity of reconstructed images.

Low-rank matrix approximation (LoRMA) techniques have emerged as powerful tools for data compression due to their ability to capture the underlying low-dimensional structure of high-resolution medical images, effectively reducing noise and preserving important features. However, this global approach often fails to capture local variations, leading to a loss of important fine details \cite{ding2018image,mcgivney2014svd}. To address this limitation, recent research \cite{wang2017hyperspectral,liu2019medical,zeng2020hyperspectral,li2024new,hamlomo2025systematic} has shifted toward patch-based and locally adaptive strategies that exploit non-local self-similarity in medical images to preserve essential anatomical features while reducing data redundancy. The fundamental idea behind these papers is to split the data matrix into overlapping patches, group structurally similar patches, and apply LoRMA for data matrix compression or denoising.

While the idea of patch-grouping is established, a critical gap remains in understanding which clustering technique is most effective at forming the homogeneous, low-variance clusters that are essential for high-fidelity, adaptive compression. Different clustering algorithms have inherent biases; some prioritize global separation, which yields good overall metrics, while others may form tighter, more homogeneous groups at the cost of creating more clusters. This trade-off directly impacts compression performance: homogeneous clusters allow for more aggressive compression (higher compression ratios) while minimizing the loss of fine details, whereas clusters with high internal variability force a conservative compression strategy to avoid introducing artifacts or losing diagnostically crucial information. This paper addresses this gap by presenting a systematic benchmarking study. In particular, we examine five clustering methods: k-means, mini-batch k-means, agglomerative hierarchical clustering, balanced iterative reducing and clustering using hierarchies and bisecting k-means, specifically on their ability to form compact, low-variance clusters suitable for adaptive LoRMA-based compression across different medical imaging modalities. We assess their effectiveness in forming compact, low-variance clusters across these modalities. These clusters are important for post-processing tasks such as per-cluster adaptive compression, since homogeneous clusters enable aggressive compression while preserving important features. We evaluated the clustering methods using the Silhouette score, Davies-Bouldin and Calinski-Harabasz index, which quantify cohesion and separation.

This work makes three contributions: we evaluated four clustering techniques using different medical imaging modalities; we used random search to tune hyperparameters for every clustering algorithm; we provide a recommendation as to which clustering technique to use on which modality.

The rest of this paper is organized as follows: Section~\ref{sec:clustering_techniques} discusses the clustering techniques used for patch grouping; Section~\ref{sec:parameter_tuning} details the hyperparameter tuning method used to optimize each clustering technique; Section~\ref{sec:evaluation_metrics} outlines the evaluation metrics used; Section~\ref{sec:results} presents experimental results and analysis; and Section~\ref{sec:conclusion} concludes with findings and future directions.
\section{Clustering techniques}\label{sec:clustering_techniques}
To exploit redundancy and structural similarities in medical images, the image $\mathbf{A} \in \mathbb{R}^{H \times W}$ is divided into a set of overlapping square patches of size $p \times p$. Each patch is extracted  by sliding a window across the image with stride $s < p$, ensuring overlap. The resulting patch $\mathbf{P}_i$ is reshaped (vectorized) into a column vector 
$\mathbf{p}_i \in \mathbb{R}^{p^2}$ by stacking its pixel intensities in a fixed 
order (e.g., row-wise). Collectively, the image yields a patch set $\boldsymbol{\mathcal{P}} = \{\mathbf{p}_1, \mathbf{p}_2, \ldots, \mathbf{p}_N\}$. Once patches are obtained, clustering algorithms are applied directly to this high-dimensional patch set to group similar patch regions. In this section, we discuss mini-batch k-means, agglomerative hierarchical clustering, balanced iterative reducing and clustering using hierarchies and bisecting k-means. The k-means theory is not discussed in this paper, and readers are referred to \cite{hamlomo2025systematic} for details.

These five clustering methods were selected since they represent both partitioning-based approaches (k-means, mini-batch k-means, bisecting k-means) and hierarchical-based approaches (agglomerative, BIRCH). This selection provides a balanced benchmark in terms of scalability, computational efficiency, and ability to capture local structure, which are critical in medical imaging tasks where both large data volumes and diagnostic fidelity must be considered.
\subsection{Mini-batch k-means}
Mini-batch k-means is a variant of the traditional k-means clustering algorithm that uses small random batches of data (mini-batches) to update the cluster centroids, rather than using the entire dataset at each iteration. The objective is to partition these patches into \( K \) clusters that minimise the within-cluster variance. Let \(\{ \boldsymbol{\mu}_1^{(0)}, \dots \boldsymbol{\mu}_K^{(0)} \} \subset \mathbb{R}^{p^2}\) denote the initial cluster centroids where each \( \boldsymbol{\mu}_k^{(0)} \in \mathbb{R}^{p^2} \) is initialized either randomly or using a heuristic such as k-means++. The clustering objective function is given by
\begin{align}
\min_{\mathcal{C}_1, \dots, \mathcal{C}_K} \sum_{k=1}^K \sum_{\mathbf{p} \in \mathcal{C}_k} \| \mathbf{p} - \boldsymbol{\mu}_k \|^2,
\end{align}
where \( \mathcal{C}_k \) denotes the set of data points assigned to cluster \(\mathcal{C}_k \). At each iteration \( t = 1, \dots, T \), a mini-batch
\begin{align}
\mathcal{B}^{(t)} = \{ \mathbf{p}_1^{(t)}, \dots, \mathbf{p}_b^{(t)} \}\subset \mathcal{P}
\end{align}
of size \( b \ll N \) is sampled uniformly at random. Each point \( \mathbf{p} \in \mathcal{B}^{(t)} \) is assigned to its nearest cluster according to
\begin{align}
k^*(\mathbf{p}) = \arg\min_{k \in \{1, \dots, K\}} \| \mathbf{p} - \boldsymbol{\mu}_k^{(t)} \|^2.
\end{align}
Let \( \mathcal{B}_k^{(t)} = \{ \mathbf{p} \in \mathcal{B}^{(t)} \mid k^*(\mathbf{p}) = k \} \) be the set of mini-batch points assigned to cluster \(\mathcal{C}_k \) at iteration \( t \). Define a learning rate for each cluster
\begin{align}
\eta_k^{(t)} = \frac{1}{N_{c_k}^{(t)} + 1},
\end{align}
where \( N_{c_k}^{(t)} \) is the cumulative number of data points assigned to cluster \(\mathcal{C}_k \) up to iteration \( t \), accounting for centroid updates at each step. The centroids are then updated using a batch-level residual correction
\begin{align}
\boldsymbol{\mu}_k^{(t+1)} = \boldsymbol{\mu}_k^{(t)} + \eta_k^{(t)} \cdot \frac{1}{|\mathcal{B}_k^{(t)}|} \sum_{\mathbf{p} \in \mathcal{B}_k^{(t)}} \left( \mathbf{p} - \boldsymbol{\mu}_k^{(t)} \right),
\end{align}
provided that \( |\mathcal{B}_k^{(t)}| > 0 \); otherwise, the centroid remains unchanged.
\subsection{Hierarchical clustering}
Hierarchical clustering groups similar data points by building a nested structure of clusters, typically represented as a dendrogram. It is broadly categorised into divisive (top-down) and agglomerative (bottom-up) methods. In this paper, we adopt agglomerative hierarchical clustering as it avoids the complexity of deciding how to split nodes and assign child levels, which are key challenges in divisive clustering \cite{ran2023comprehensive}. In agglomerative hierarchical clustering, the process begins at iteration $t=0$ with 
\begin{align}
\boldsymbol{\mathcal{C}^{(0)}} = \bigl\{\mathcal{C}_1^{(0)}, \dots, \mathcal{C}_N^{(0)}\bigr\},
\quad \mathcal{C}_i^{(0)} = \{\mathbf{p}_i\}.
\end{align}
A linkage-based distance between any two clusters \(\mathcal{C}_i\) and \(\mathcal{C}_j\) is then defined. For instance, using single linkage
\begin{align}
d_{\mathrm{single}}\bigl(\mathcal{C}_i,\mathcal{C}_j\bigr) 
= \min_{\mathbf{x}\in \mathcal{C}_i,\, \mathbf{y}\in \mathcal{C}_j}\|\mathbf{x}-\mathbf{y}\|,
\end{align}
complete linkage
\begin{align}
d_{\mathrm{complete}}\bigl(\mathcal{C}_i,\mathcal{C}_j\bigr) 
= \max_{\mathbf{x}\in \mathcal{C}_i,\, \mathbf{y}\in \mathcal{C}_j}\|\mathbf{x}-\mathbf{y}\|,
\end{align}
or average linkage
\begin{align}
d_{\mathrm{avg}}\bigl(\mathcal{C}_i,\mathcal{C}_j\bigr) 
= \frac{1}{|\mathcal{C}_i|\;|\mathcal{C}_j|} 
\sum_{\mathbf{x}\in \mathcal{C}_i}\sum_{\mathbf{y}\in \mathcal{C}_j}\|\mathbf{x}-\mathbf{y}\|.
\end{align}
At each iteration \(t\), the pair \(\bigl(a,b\bigr)\) minimizing the chosen linkage distance is identified
\begin{align}
(a,b) = \arg\min_{i\neq j}\;d\bigl(\mathcal{C}_i^{(t)},\mathcal{C}_j^{(t)}\bigr).
\end{align}
These two clusters are merged into a new cluster
\begin{align}
\mathcal{C}_{\text{new}}^{(t+1)} 
= \mathcal{C}_a^{(t)} \cup \mathcal{C}_b^{(t)},
\end{align}
and the cluster set is updated as follows:
\begin{align}
\mathcal{C}^{(t+1)} 
= \bigl(\mathcal{C}^{(t)} \setminus \{\mathcal{C}_a^{(t)}, \mathcal{C}_b^{(t)}\}\bigr)\;\cup\;\{\mathcal{C}_{\text{new}}^{(t+1)}\}.
\end{align}
This process continues until a single cluster remains, yielding a dendrogram whose merge heights record the linkage distances. Alternatively, using Ward’s criterion, one merges at each step the pair \(\bigl(a,b\bigr)\) that minimises the increase in total within-cluster variance.  If \(\boldsymbol{\mu}_i^{(t)}\) denotes the centroid of \(\mathcal{C}_i^{(t)}\), then merging \(\mathcal{C}_a\) and \(\mathcal{C}_b\) incurs
\begin{align}
\Delta E = \frac{|\mathcal{C}_a^{(t)}|\,|\mathcal{C}_b^{(t)}|}{|\mathcal{C}_a^{(t)}| + |\mathcal{C}_b^{(t)}|} \,\bigl\|\boldsymbol{\mu}_a^{(t)} - \boldsymbol{\mu}_b^{(t)}\bigr\|_2^2,
\end{align}
and the merge yielding the minimum $\Delta E$ is chosen. Once the full hierarchy is built, a flat partition into \(K\) clusters is obtained by “cutting” the dendrogram at the appropriate linkage threshold, producing clusters \(\mathcal{C}_1, \dots, \mathcal{C}_K\).  
\subsection{Balanced iterative reducing and clustering using hierarchies}
Balanced iterative reducing and clustering using hierarchies (BIRCH) is a hierarchical clustering algorithm designed to efficiently handle large datasets by incrementally constructing a compact summary structure known as a clustering feature (CF) tree. Each node of the CF-tree stores a clustering feature
\begin{align}
\mathrm{CF}(\mathcal{C}) = \bigl(N_{\mathcal{C}},\,\mathbf{LS}_{\mathcal{C}},\,SS_{\mathcal{C}}\bigr),
\end{align}
where \(N_{\mathcal{C}}=|\mathcal{C}|\) counts its patches, \(\mathbf{LS}_{\mathcal{C}}=\sum_{\mathbf{p}\in\mathcal{C}}\mathbf{p}\) is their vector sum, and \(SS_{\mathcal{C}}=\sum_{\mathbf{p}\in\mathcal{C}}\|\mathbf{p}\|^2\) records their squared-norm total. These three quantities add entry-wise over any disjoint union, so that a parent node’s CF exactly summarises the union of its children’s patches. That is, if \(\mathcal{C} = \mathcal{C}_a \cup \mathcal{C}_b\) and \(\mathcal{C}_a \cap \mathcal{C}_b = \emptyset\), then
\begin{align}
\mathrm{CF}(\mathcal{C}) = \mathrm{CF}(\mathcal{C}_a) + \mathrm{CF}(\mathcal{C}_b).
\end{align}
This ensures that every node’s summary perfectly captures all the patches it contains. From any CF, one computes the centroid \(\mathbf{LS}_{\mathcal{C}}/N_{\mathcal{C}}\) and radius
\begin{align}
r(\mathcal{C}) = \sqrt{\frac{SS_{\mathcal{C}}}{N_{\mathcal{C}}} - \Bigl\|\frac{\mathbf{LS}_{\mathcal{C}}}{N_{\mathcal{C}}}\Bigr\|^2},
\end{align}
which quantifies the root-mean-square (RMS) distance of points from the centroid, reflecting the cluster’s spread. The CF-tree is controlled by a branching factor \(B\) (maximum entries per node), which limits the number of entries per node, and a threshold \(T\), which caps the allowable radius of any leaf-level feature. Patches are gradually incorporated by always choosing, at each internal node, the entry whose feature would grow least in squared radius when augmented by the new patch, that is
\begin{align}
j^* \;=\;\arg\min_{1\le j\le B}\Bigl[r\bigl(\mathrm{CF}_j + \mathbf{p}\bigr)^2 \;-\; r(\mathrm{CF}_j)^2\Bigr].
\end{align}
The selected feature then increases its counts and sums by those of \(\mathbf{p}\), and the process continues down to a leaf. If the updated radius of that leaf feature remains at most \(T\), the patch merges into it; otherwise, a fresh feature consisting of \(\bigl(1,\mathbf{p},\|\mathbf{p}\|^2\bigr)\) is introduced. Whenever a node’s number of features exceeds \(B\), it is split by identifying two “seed” features whose centroids lie furthest apart and partitioning the remainder according to the nearest seed. Once all \(N\) patches are summarised in the leaf features, their centroids are given by
\begin{align}
\{\mathbf{LS}_\ell / N_\ell\}_{\ell=1}^L,
\end{align}
where \(L\) denotes the total number of leaf-level CF entries. These centroids provide a compact representation of \(\mathcal{P}\), to which a final flat clustering method (such as k-means or agglomerative clustering) is applied in order to produce the desired clusters \(\boldsymbol{\mathcal{C}}\).
\subsection{Bisecting k-means}
Bisecting k-means is a hierarchical clustering algorithm that recursively partitions a patch set \(\mathcal{P}\) by applying two-means clustering to selected subsets until \(K\) clusters are formed. It begins with a single cluster \(\boldsymbol{\mathcal{C}}^{(0)} = \{ \mathcal{P} \}\), and at each iteration \(t\), a cluster \(\mathcal{C}_{\text{sel}}^{(t)} \in \boldsymbol{\mathcal{C}}^{(t)}\) is chosen for bisection. A two-means clustering is applied to \(\mathcal{C}_{\text{sel}}^{(t)}\), yielding two sub-clusters \(\mathcal{C}_a^{(t+1)}\) and \(\mathcal{C}_b^{(t+1)}\), such that
\begin{align}
\mathcal{C}_{\text{sel}}^{(t)} = \mathcal{C}_a^{(t+1)} \cup \mathcal{C}_b^{(t+1)}
\quad \text{and} \quad
\mathcal{C}_a^{(t+1)} \cap \mathcal{C}_b^{(t+1)} = \emptyset,
\end{align}
with corresponding centroids \(\boldsymbol{\mu}_a^{(t)}, \boldsymbol{\mu}_b^{(t)} \in \mathbb{R}^{p^2}\). The split quality is measured by the within-cluster sum of squares, that is
\begin{align}
\mathrm{WCSS}^{(t)} 
= \sum_{\mathbf{p} \in \mathcal{C}_a^{(t+1)}} \|\mathbf{p} - \boldsymbol{\mu}_a^{(t)}\|^2
+ \sum_{\mathbf{p} \in \mathcal{C}_b^{(t+1)}} \|\mathbf{p} - \boldsymbol{\mu}_b^{(t)}\|^2.
\end{align}
To improve stability, the bisection is repeated \(n_{\text{try}}\) times with different initializations, and the split minimizing \(\mathrm{WCSS}^{(t)}\) is chosen. The updated clustering becomes
\begin{align}
\boldsymbol{\mathcal{C}}^{(t+1)} = \left(\boldsymbol{\mathcal{C}}^{(t)} \setminus \left\{ \mathcal{C}_{\text{sel}}^{(t)} \right\} \right) \cup \left\{ \mathcal{C}_a^{(t+1)},\, \mathcal{C}_b^{(t+1)} \right\}.
\end{align}
This process continues until \(|\boldsymbol{\mathcal{C}}^{(t)}| = K\), producing a final clustering \(\boldsymbol{\mathcal{C}} = \{ \mathcal{C}_1, \dots, \mathcal{C}_K \}\).
\section{Hyperparameter Tuning}\label{sec:parameter_tuning}
Hyperparameter tuning aims to identify the set of model parameters that yields the best performance on a given task. In clustering, hyperparameters such as the number of clusters, batch sizes, linkage criteria, and tree thresholds can greatly influence the quality of the resulting clusters. To explore the hyperparameter space, the configuration vector is expressed as
\begin{align}
\boldsymbol{\theta} = \bigl(K,\alpha_{1},\dots,\alpha_{d}\bigr)\;\in\;\boldsymbol{\Theta},
\end{align}
where \(\boldsymbol{\Theta}\) denotes the feasible hyperparameter space. This space is constructed as a Cartesian product of sets, that is  
\begin{align}
\boldsymbol{\Theta} = \mathcal{K}\times \mathcal{A}_{1}\times \cdots \times \mathcal{A}_{d},
\quad
\mathcal{K} = \{K_{\min},\dots,K_{\max}\},
\end{align}
where each \(\mathcal{A}_{j}\) denotes the admissible set for hyperparameter \(\alpha_{j}\) and \(\mathcal{K}\) denotes the range of possible values for the number of clusters. To optimise, we perform a random search with budget \(T\) by drawing
\begin{align}
\boldsymbol{\theta}_{1},\cdots,\boldsymbol{\theta}_{T}
\;\overset{\mathrm{i.i.d}}{\sim}\;\mathcal{U}(\boldsymbol{\Theta}).
\end{align}
Let
\begin{align}
\Gamma_{\boldsymbol{\theta}}\colon \mathcal{P}
\;\longrightarrow\;\boldsymbol{\mathcal{C}}(\mathcal{P}),
\end{align}
denote the clustering operator where \(\boldsymbol{\mathcal{C}}(\mathcal{P})\) is the set of all partitions of \(\mathcal{P}\). For fixed \(\boldsymbol{\theta}\),
\begin{align}
\boldsymbol{\mathcal{C}}_{\boldsymbol{\theta}}
= \Gamma_{\boldsymbol{\theta}}(\mathcal{P})
= \bigl\{\mathcal{C}_{1}(\boldsymbol{\theta}),\dots,\mathcal{C}_{K}(\boldsymbol{\theta})\bigr\},
\end{align}
where each \(\mathcal{C}_{k}(\boldsymbol{\theta})\subseteq\mathcal{P}\) and
\begin{align}
\bigcup_{k=1}^{K}\mathcal{C}_{k}(\boldsymbol{\theta})
=\mathcal{P},
\quad
\mathcal{C}_{i}(\boldsymbol{\theta})\cap\mathcal{C}_{j}(\boldsymbol{\theta})
=\emptyset
\quad(i\neq j).
\end{align}
For each draw \(\boldsymbol{\theta}_{t}\), we apply \(\Gamma_{\boldsymbol{\theta}_{t}}\) to \(\mathcal{P}\) to obtain the partition \(\{\mathcal{C}_{k}(\boldsymbol{\theta}_{t})\}_{k=1}^{K}\) and compute its average Silhouette score
\begin{align}
f_{t}
=\frac{1}{N}\sum_{i=1}^{N}
\frac{b(i;\boldsymbol{\theta}_{t})-a(i;\boldsymbol{\theta}_{t})}
{\max\{\,a(i;\boldsymbol{\theta}_{t}),\,b(i;\boldsymbol{\theta}_{t})\}}.
\end{align}
We record the pairs \(\bigl(\boldsymbol{\theta}_{t},f_{t}\bigr)\) for \(t=1,\dots,T\) and select
\begin{align}
\boldsymbol{\theta}^{*}
=\arg\max_{t\in\{1, 2, \dots, T\}}f_{t}.
\end{align}
\section{Evaluation Metrics}\label{sec:evaluation_metrics} 
The cluster quality metrics Silhouette score, Calinski-Harabasz (CH) index and Davies-Bouldin (DB) index were used to evaluate the performance of each clustering technique. These evaluation metrics assess the quality of clustering results by measuring cohesion (how compact each cluster is) and separation (how well clusters are distinguished from one another). The Silhouette score quantifies how well each data point \(\mathbf{p}_i\) lies within its assigned cluster compared to other clusters. For a patch \(\mathbf{p}_i\), let \(a(i)\) denote the average distance from \(\mathbf{p}_i\) to all other points in the same cluster (intra-cluster distance), that is
\begin{align}
a(i)
\;=\;
\frac{1}{\bigl|\mathcal{C}_k\bigr| - 1}
\sum_{\substack{\mathbf{p}_j \in \mathcal{C}_k \\ j \neq i}}
\bigl\lVert \mathbf{p}_i - \mathbf{p}_j \bigr\rVert_2
\quad\text{for }\mathbf{p}_i\in\mathcal{C}_k
\end{align}
and let \(b(i)\) be the minimum average distance from \(\mathbf{p}_i\) to all points in the nearest different cluster (inter-cluster distance), that is
\begin{align}
b(i)
\;=\;
\min_{m \neq k}
\;\frac{1}{\bigl|\mathcal{C}_m\bigr|}
\sum_{\mathbf{p}_j \in \mathcal{C}_m}
\bigl\lVert \mathbf{p}_i - \mathbf{p}_j \bigr\rVert_2.
\end{align}
Therefore, the silhouette value \(s(i)\) is defined as
\begin{align}
s(i) = \frac{b(i) - a(i)}{\max\{a(i),\,b(i)\}},
\end{align}
and lies in the interval \([-1, 1]\). A value near \(+1\) indicates that \(\mathbf{p}_i\) is well matched to its own cluster and poorly matched to others. A value near \(0\) suggests it lies on the decision boundary between two clusters, while negative values indicate likely misclassifications. The overall silhouette score is the average of \(s(i)\) over all patches.

On the other hand, the CH index, also known as the variance ratio criterion, measures the ratio of between-cluster dispersion to within-cluster dispersion. Let \(\mathbf{c}_k\) be the centroid of cluster \(\mathcal{C}_k\), and let \(\mathbf{c}\) be the global centroid of all data. The CH index is defined as
\begin{align}
\mathrm{CH} = \frac{\mathrm{Tr}(\mathbf{B}_K)}{\mathrm{Tr}(\mathbf{W}_K)} \cdot \frac{N - K}{K - 1},
\end{align}
where \(\mathrm{Tr}(\mathbf{B}_K)\) denotes the trace of the between-cluster dispersion matrix
\begin{align}
\mathrm{Tr}(\mathbf{B}_K) = \sum_{k=1}^K N_k \|\mathbf{c}_k - \mathbf{c}\|^2,
\end{align}
and \(\mathrm{Tr}(\mathbf{W}_K)\) denotes the trace of the within-cluster dispersion matrix
\begin{align}
\mathrm{Tr}(\mathbf{W}_K) = \sum_{k=1}^K \sum_{\mathbf{p}_i \in \mathcal{C}_k} \|\mathbf{p}_i - \mathbf{c}_k\|^2.
\end{align}
Higher values of \(\mathrm{CH}\) indicate more distinct and well-separated clusters.

The DB index quantifies the average similarity between each cluster and its most similar one, with similarity based on intra-cluster distances and inter-cluster separation. For each cluster \(\mathcal{C}_k\), define
\begin{align}
s_k = \frac{1}{|\mathcal{C}_k|} \sum_{\mathbf{p}_i \in \mathcal{C}_k} \|\mathbf{p}_i - \mathbf{c}_k\|,
\end{align}
the average distance of points to the cluster centroid. Then for clusters \(\mathcal{C}_k\) and \(\mathcal{C}_m\), define
\begin{align}
R_{k,m} = \frac{s_k + s_m}{\|\mathbf{c}_k - \mathbf{c}_m\|},
\end{align}
where \(R_{k,m}\) denotes the similarity measure between \(\mathcal{C}_k\) and \(\mathcal{C}_m\). The DB index is the average over the maximum \(R_{k,m}\) for each cluster
\begin{align}
\mathrm{DB} = \frac{1}{K} \sum_{k=1}^K \max_{m\ne k} R_{k,m}.
\end{align}
Lower values of \(\mathrm{DB}\) indicate better clustering quality, with compact, well-separated clusters.
\section{Results and Discussion}\label{sec:results}
This study uses three publicly available medical imaging datasets sourced from Kaggle. From each dataset, a single representative image was randomly selected. The brain MRI was sourced from the \href{https://www.kaggle.com/datasets/awsaf49/brats20-dataset-training-validation}{BraTS2020 dataset}, which contains multi-institutional, preoperative scans at 1 mm\(^3\) resolution (including T1, T1Gd, T2, and FLAIR sequences) with expert-annotated segmentation masks. The breast ultrasound image came from the \href{https://www.kaggle.com/datasets/aryashah2k/breast-ultrasound-images-dataset}{Breast Ultrasound Images dataset}, comprising 780 images labelled as normal, benign, or malignant, each with a lesion segmentation mask. Finally, the chest X-ray was selected from the \href{https://www.kaggle.com/datasets/anasmohammedtahir/covidqu}{COVIDqu dataset}, a collection of 33,920 images categorised as COVID-19, non-COVID pneumonia, or normal, all accompanied by lung segmentation masks. 

\begin{figure}[ht]
    \centering
    \includegraphics[width=\textwidth]{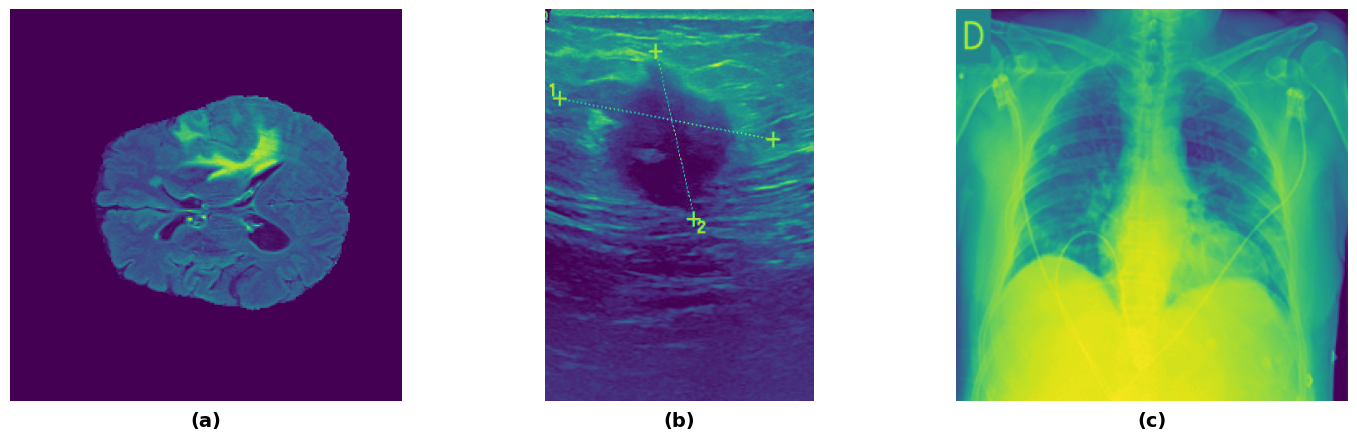}
    \caption{Shows three medical imaging modalities analysed in this study: (a) a brain MRI scan, (b) an ultrasound image annotated with calliper measurements, and (c) a frontal chest X-ray radiograph.}
    \label{fig:original_images}
\end{figure}

Fig.~\ref{fig:original_images} shows three medical imaging modalities analysed in this study: (a) a brain MRI scan, (b) an ultrasound image annotated with calliper measurements, and (c) a frontal chest X-ray radiograph. All images were split into overlapping patches of a fixed size, and clustering methods were applied to these patches to evaluate how well each technique groups similar regions across modalities. Hyperparameters for each clustering technique (e.g., number of clusters for k-means variants, linkage criteria for hierarchical clustering, threshold/branching factor for BIRCH) were tuned using random search over a predefined range to ensure each method achieved optimal results.

Table~\ref {tab:clustering_metrics} summarises clustering performance metrics: Silhouette score, DB index, CH index, and the optimal number of clusters selected for each method and modality. 
\begin{table}[ht]
  \centering
  \caption{Comparison of clustering performance metrics across different medical imaging modalities (MRI, ultrasound, X-ray) using various clustering methods. The performance evaluation metrics include Silhouette score, DB and CH index. We also record the optimal number of clusters for each clustering technique.}
  \label{tab:clustering_metrics}
  \begin{adjustbox}{max width=\textwidth}
  \begin{tabular}{@{}l
    *{4}{c}
    *{4}{c}
    *{4}{c}@{}}
    \toprule
    \textbf{Method}
      & \multicolumn{4}{c}{\textbf{MRI}}
      & \multicolumn{4}{c}{\textbf{Ultrasound}}
      & \multicolumn{4}{c}{\textbf{X-ray}} \\
    \cmidrule(lr){2-5} \cmidrule(lr){6-9} \cmidrule(lr){10-13}
      & \makecell{Silhouette}
      & \makecell{Davies-\\Bouldin}
      & \makecell{Calinski-\\Harabasz}
      & \makecell{No. of\\clusters}
      & \makecell{Silhouette}
      & \makecell{Davies-\\Bouldin}
      & \makecell{Calinski-\\Harabasz}
      & \makecell{No. of\\clusters}
      & \makecell{Silhouette}
      & \makecell{Davies-\\Bouldin}
      & \makecell{Calinski-\\Harabasz}
      & \makecell{No. of\\clusters} \\
    \midrule
    K-means           
      & 0.814 & 0.705 & 8196.349 & 3  
      & 0.594 & 0.589 & 14449.452 & 2 
      & 0.540 & 0.631 & 7576.080 & 2 \\
    Mini-batch k-means  
      & 0.779 & 0.963 & 6311.616  & 4   
      & 0.318 & 1.015 & 9395.139 & 4  
      & 0.456 & 0.714 & 8967.025 & 4  \\
    Agglomerative    
      & 0.801 & 0.789 & 2785.943  & 7   
      & 0.511 & 0.485 & 920.711   & 11  
      & 0.444 & 0.902 & 3732.178  & 7 \\
    Birch            
      & 0.746 & 1.366 & 4019.569  & 6   
      & 0.354 & 1.169 & 8465.552  & 4  
      & 0.420 & 0.759 & 7654.140  & 4 \\
    Bisecting K-means             
      & 0.808 & 0.425 & 9421.159  & 2   
      & 0.594 & 0.589 & 14449.296 & 2  
      & 0.540 & 0.631 & 7575.771  & 2 \\
    \bottomrule
  \end{tabular}
  \end{adjustbox}
\end{table}

For the MRI modality, standard k-means achieves the highest Silhouette score (0.814), followed closely by bisecting k-means (0.808) and agglomerative clustering (0.801). The lowest DB index is observed for bisecting k-means (0.425), indicating the most compact and well-separated clusters. It also attains the highest CH score (9421.16), showing strong clustering performance. While mini-batch k-means produces a reasonably high Silhouette score (0.779), its DB index is the highest (0.963), suggesting less optimal compactness compared to the other methods. For the ultrasound modality, standard k-means and bisecting k-means tie for the highest Silhouette score (0.594) and lowest DB index (0.589), with very similar CH scores (14449.45 and 14449.30, respectively), making them the top-performing methods. Agglomerative clustering, though exhibiting a reasonably strong Silhouette score (0.511) and the lowest DB index (0.485), results in a significantly lower CH score (920.71), which may reflect the penalty imposed by the CH index on the higher number of clusters (11) and potential imbalance in cluster sizes. In the X-ray modality, standard k-means and bisecting k-means both achieve the highest Silhouette score (0.540) and the lowest DB index (0.631), with nearly identical CH scores (7576.08 and 7575.77, respectively). Although mini-batch k-means attains a competitive CH score (8967.03), its lower Silhouette score (0.456) and higher DB index (0.714) suggest weaker intra-cluster cohesion and separation.

\begin{figure}
    \centering
    \includegraphics[width=\textwidth]{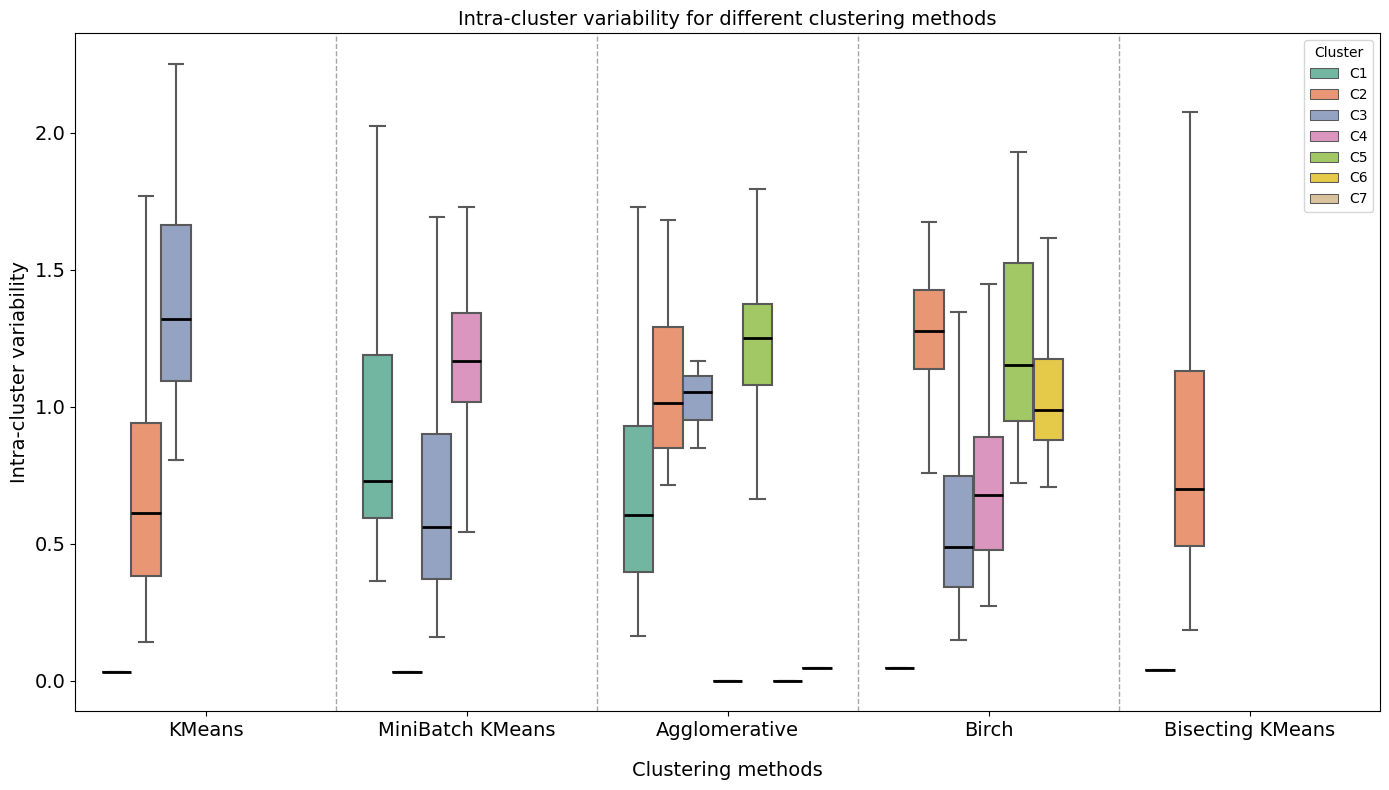}
    \caption{Intra-cluster variability for MRI patches clustered using k-means, mini-batch k-means, agglomerative, birch and bisecting k-means methods. Each box-plot represents the variability within a specific cluster (C1–C7), with lower values indicating more homogeneous clusters.}
    \label{fig:mri_cluster_variability}
\end{figure}

The results of this study show that standard k-means and bisecting k-means generally yield strong clustering performance based on Silhouette scores, DB and CH indices, indicating good cohesion and separation among clusters. However, when a post-processing task such as per-cluster adaptive compression is considered, intra-cluster variability becomes crucial. For example, Fig.~\ref{fig:mri_cluster_variability} shows a box-plot comparison of intra-cluster variability for different clustering methods applied to MRI image patches. Each box-plot represents the distribution of intra-cluster variability for a specific cluster across different clustering methods. As shown in Fig.~\ref{fig:mri_cluster_variability}, while k-means and bisecting k-means perform well overall in terms of clustering metrics, they form a small number of clusters with relatively high intra-cluster variability, which limits the flexibility of adaptive cluster-wise compression since aggressive compression within these clusters may result in the loss of important or clinically relevant features. While agglomerative clustering is more computationally intensive, it forms tighter, more homogeneous clusters with low intra-cluster variability, yielding low-rank structures that enable more effective compression while minimising the risk of degrading clinically relevant information.

\begin{figure}
    \centering
    \includegraphics[width=\textwidth]{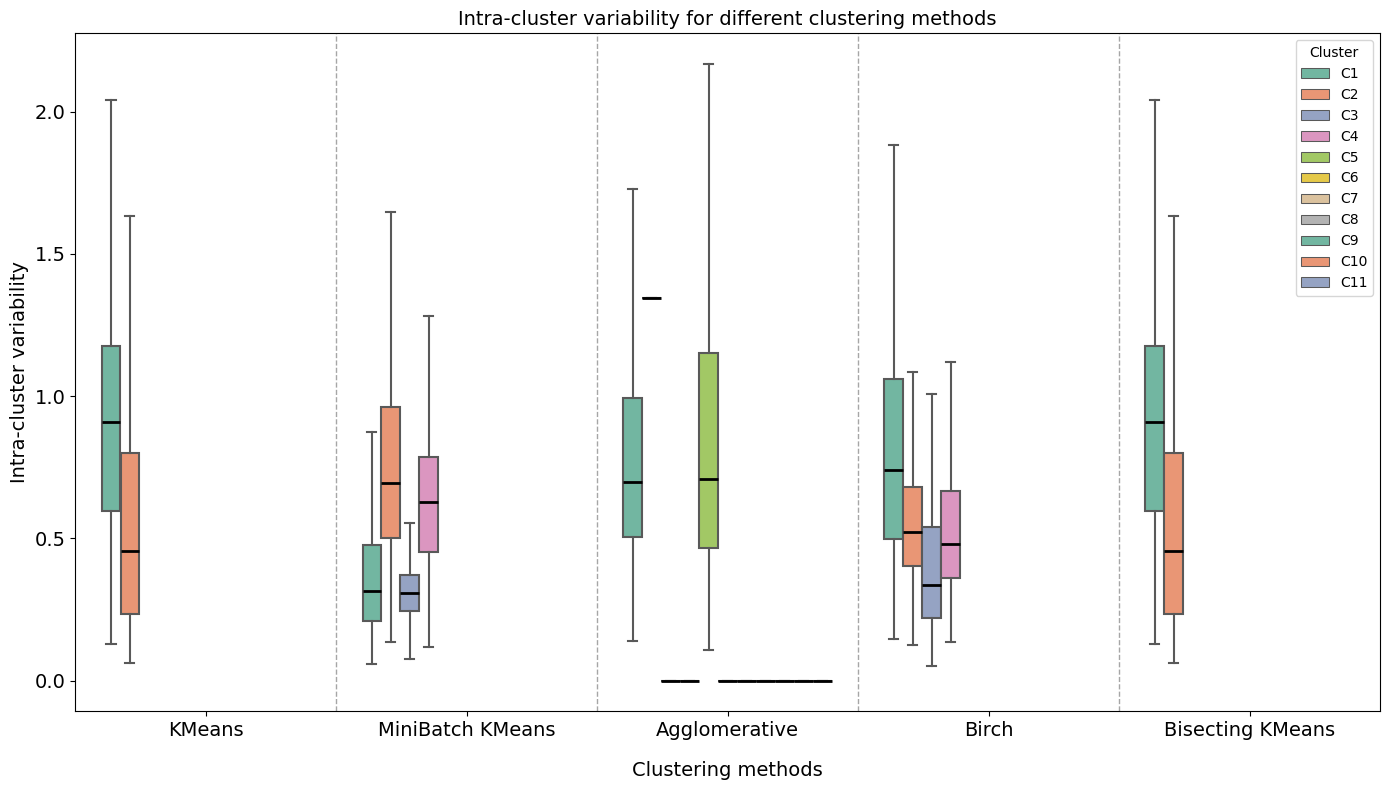}
    \caption{Intra-cluster variability for Ultrasound patches clustered using k-means, mini-batch k-means, agglomerative, birch and bisecting k-means methods. Each box-plot represents the variability within a specific cluster (C1–C11), with lower values indicating more homogeneous clusters.}
    \label{fig:ultrasound_cluster_variability}
\end{figure}

On the other hand, Fig.~\ref{fig:ultrasound_cluster_variability} indicates that k-means and bisecting k-means produce two clusters with high intra-cluster variability, which limits their suitability for adaptive compression. Although agglomerative clustering performs slightly less in terms of clustering metrics, it forms a larger number of more homogeneous clusters. This makes it more suitable for adaptive cluster-wise compression in ultrasound images, as it enables more effective low-rank approximations while preserving essential diagnostic details.

\begin{figure*}
    \centering
    \includegraphics[width=0.9\textwidth]{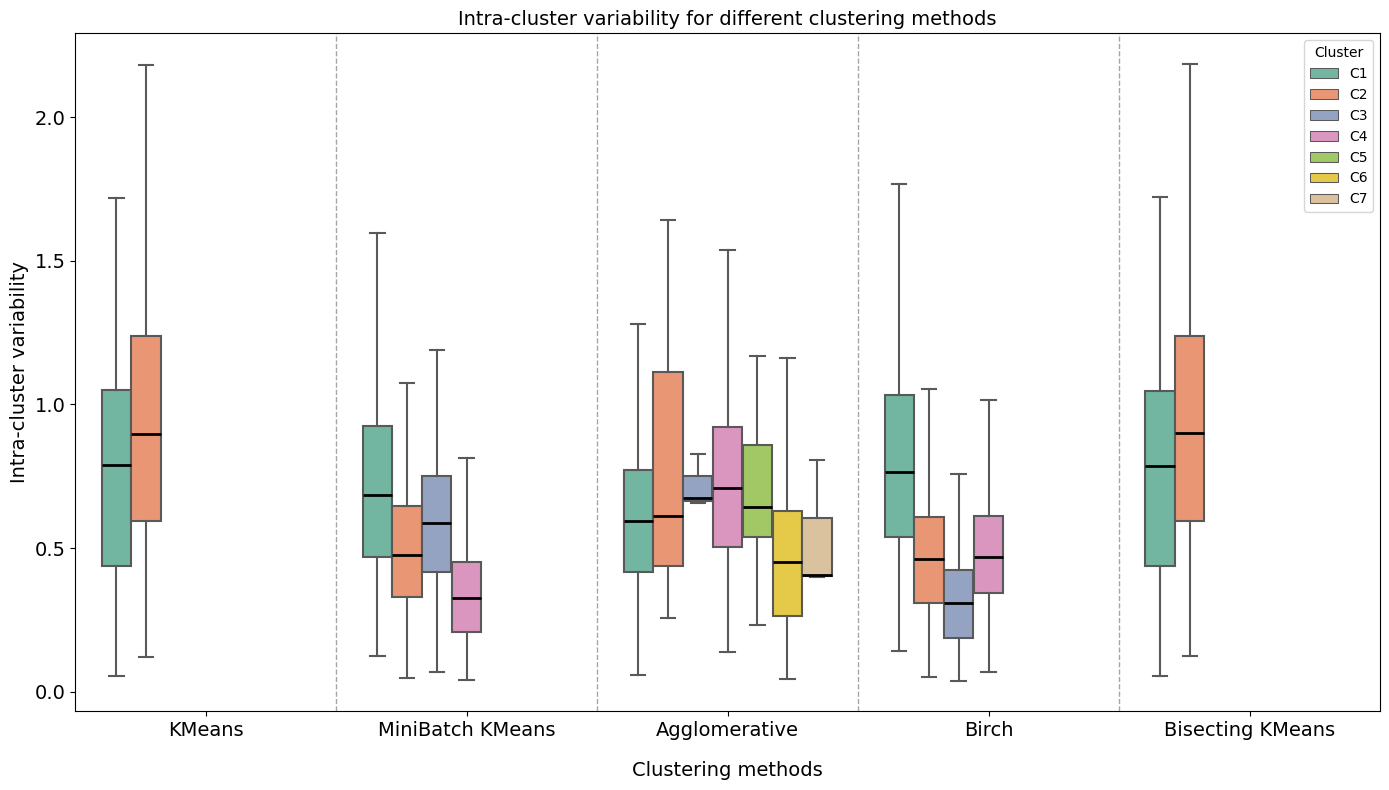}
    \caption{Intra-cluster variability for X-ray patches clustered using k-means, mini-batch k-means, agglomerative, and Birch methods. Each box-plot represents the variability within a specific cluster (C1–C7), with lower values indicating more homogeneous clusters.}
    \label{fig:xray_cluster_variability}
\end{figure*}

While mini-batch k-means performs slightly below k-means and bisecting k-means in standard clustering metrics, Fig.~\ref{fig:xray_cluster_variability} shows that it generates a larger number of clusters with lower intra-cluster variability, making it more effective for adaptive cluster-wise compression.
\section{Conclusion}\label{sec:conclusion}
This study explored the effectiveness of clustering-based techniques for grouping structurally similar patches in high-resolution medical images, enabling adaptive LoRMA-based compression during post-processing. The primary focus of our work is to investigate how the choice of clustering algorithm impacts not just standard clustering metrics, but more importantly, the intra-cluster homogeneity that is critical for post-processing tasks like compression. By comparing standard k-means, mini-batch k-means, agglomerative hierarchical clustering, BIRCH and bisecting k-means across MRI, ultrasound, and chest X-ray modalities, we evaluated clustering performance using Silhouette scores, DB index, and CH index. Our results indicate that bisecting k-means and standard k-means generally achieved strong cluster quality in terms of cohesion and separation, particularly in the MRI modality, as reflected by high Silhouette and CH scores. However, as shown in Figs.~\ref{fig:mri_cluster_variability} and \ref{fig:ultrasound_cluster_variability}, these methods form a small number of clusters with high intra-cluster variability, limiting their suitability for adaptive per-cluster compression. In contrast, agglomerative clustering, while more computationally intensive, produced tighter and more homogeneous clusters with lower intra-cluster variability for both MRI and ultrasound image patches, making it more appropriate for compression in those modalities. For chest X-ray images, mini-batch k-means achieved the best balance between cluster quality and intra-cluster compactness, suggesting it is well-suited for adaptive compression in that modality. Meanwhile, BIRCH showed relatively lower performance across all modalities in both clustering quality and intra-cluster variability, making it the least suitable for the compression framework considered.

These findings suggest that the choice of clustering technique should be guided by both the imaging modality and the trade-off between computational efficiency and compression fidelity. Mini-batch k-means provides an efficient and effective option for chest X-ray images, where clustering quality and compactness align well. However, for MRI and ultrasound images, agglomerative clustering is more suitable due to its ability to form tighter, more homogeneous clusters that preserve fine structural and clinical details, as evidenced by lower intra-cluster variability. In future work, we will examine how variations in patch size influence clustering behaviour and compression performance across different medical imaging modalities.
\subsubsection{Acknowledgments}
This research was supported in part by the Department of Higher Education and Training (DHET) through the University Staff Doctoral Programme (USDP) and in part by the National Research Foundation of South Africa (Ref No.CSRP23040990793). The authors would like to thank Rhodes University for their financial support. 
%
%
%
\bibliographystyle{unsrt}
\bibliography{bibliography}
\end{document}